\documentclass[aps,prl,floatfix,superscriptaddress,twocolumn,footinbib]{revtex4-1}
\usepackage{graphicx}% Include figure files
\usepackage{dcolumn}% Align table columns on decimal point
\usepackage{bm}% bold math
\usepackage{makecell}
\usepackage{color}
\usepackage{amsmath}
\usepackage{latexsym}
\usepackage{amssymb}
\usepackage{graphics,epstopdf}
\usepackage{hyperref}
\usepackage{xcolor}
\hypersetup{
    colorlinks = true,
    linkcolor =blue,
	citecolor=blue,
	urlcolor=blue
}

\newcommand{\ed}{\end{document}}
\newcommand{\beq}{\begin{equation}}
\newcommand{\eeq}{\end{equation}}

\usepackage{mathtools}
\usepackage{comment}
\usepackage[T1]{fontenc}

\begin{document}

\title{Extended No-Go Theorem and Its Verification in Spin-Selective Transport through Chiral Cyclodextrin Assemblies}

\author{Kai-Yuan Zhang}
 \affiliation{School of Physics and Wuhan National High Magnetic Field Center,
Huazhong University of Science and Technology, Wuhan 430074, People's Republic of China.}
\author{Shu-Zheng Zhou}
 \affiliation{School of Physics and Wuhan National High Magnetic Field Center,
Huazhong University of Science and Technology, Wuhan 430074, People's Republic of China.}
\author{Xi Sun}
 \affiliation{School of Physics and Wuhan National High Magnetic Field Center,
Huazhong University of Science and Technology, Wuhan 430074, People's Republic of China.}
\author{Hua-Hua Fu}
\altaffiliation{Corresponding author.\\ hhfu@hust.edu.cn}%Lines break automatically or can be forced with \\
\affiliation{School of Physics and Wuhan National High Magnetic Field Center,
Huazhong University of Science and Technology, Wuhan 430074, People's Republic of China.}
\affiliation{Institute for Quantum Science and Engineering, Huazhong University of Science and Technology, Wuhan, Hubei 430074, China.}

\date{\today}

\begin{abstract}
The classical two-terminal single-channel no-go theorem relies on time-reversal symmetry (TRS) and scattering-matrix unitarity. In open systems, omitted channels or environment leakage break unitarity, so TRS and the symplectic reciprocity symmetry TRS\(^{\dagger}\) are no longer equivalent, invalidating the classical premises. We prove that if the two-terminal single-channel scattering matrix satisfies TRS\(^{\dagger}\) plus central inversion symmetry, transmission remains no spin selectivity; this result provides an important complement to the classical no-go theorem in nonunitary regime. When these conditions are violated or the single-channel constraint is lifted, spin-selective transport occurs. In charge transport through devices based on chiral \(\gamma\)-cyclodextrin molecular assemblies, we verify separately the classical no-go theorem in unitary case and its supplementary version together with its multichannel corollary in nonunitary case: when the conditions for the corresponding no-go theorem hold, the chirality-induced spin selectivity (CISS) of the assembly disappears, whereas when they are violated, CISS emerges. These results extend the no-go theorem to nonunitary scattering systems and establish an effective theoretical framework for studying CISS in real molecular assemblies.
\end{abstract}
\maketitle

Generating observable and controllable spin-selective transport in nonmagnetic materials without external magnetic fields remains a key challenge in spintronics~\cite{Eto2005,Gelabert2010,Santos2016,VanRuitenbeek2023}. Spin-orbit coupling (SOC) offers a viable mechanism, yet its capability to produce net transmitted spin polarization (SP) is constrained by device symmetry and lead channel structure~\cite{Dednam2023,Zollner2020,Chico2015,Zhang2017}. The classical no-go theorem states that in a two-terminal system with time-reversal symmetry (TRS) and a unitary scattering matrix, unpolarized injection yields zero SP if each terminal has only one independent transport channel, as illustrated in Figs.~1(a) and 1(b)~\cite{Zhai2005,Kiselev2005,Krich2008,Bardarson2008,Jacquod2013}. In real devices, however, coupling to the environment or omitted ports is unavoidable~\cite{Klein2023,Smorka2025}. This leakage renders the two-terminal scattering nonunitary, thereby invalidating the premise of the classical theorem~\cite{Matityahu2016,Aharony2025}. This naturally raises a critical question: in these open and nonunitary scattering systems, do other symmetry conditions exist that can enforce strictly zero SP in electron transport?

Answering this question requires distinguishing the symmetry contitions of the whole system from the constraints satisfied by the reduced scattering matrix. A whole system with TRS and unitarity satisfies a symplectic reciprocity constraint associated with $\mathrm{TRS}^{\dagger}$, which is equivalent to TRS only for unitary scattering~\cite{Mello1991,Kawabata2019,KawabataRyu2021}. While the reduced matrix inherits TRS if the observed channels contain complete time-reversal partners, leakage into unobserved channels breaks this constraint. Consequently, nonmagnetic open devices can occupy symmetry regions beyond the classical no-go theorem, such as regions (II) and (IV) in Fig.~1. Here, we show that when the reduced scattering satisfies TRS$^\dagger$ and the whole system preserves the central inversion symmetry $\mathcal{I}$ exchanging the two terminals, the single-channel (spin excluded) transmission matrix must be a spin scalar. Thus, even for nonunitary observed scattering, unpolarized injection yields strictly SP=0. We refer to this protection, covering regions (I) and (II), as an important complement to the classical no-go theorem as described in Fig.~1(c).

To test this theorem, we construct devices from \(\gamma\)-cyclodextrin molecules~\cite{CyclodextrinReview2022,Wu2024MirrorCyclodextrins,MirrorCyclodextrinActions2025}. The central scattering region consists of an enantiomer pair coupled at their large rims; adjusting electrode contacts~\cite{Guo2014} and relative molecular orientation controls the central inversion symmetry \(\mathcal{I}\) of the full device. A discrete geometric SOC describes the spin-dependent electron dynamics, while electrode coupling and environmental leakage are explicitly included. In the single-channel regime, when the device lies in regions (I) and (II) of Fig.~1(c), transmitted SP is strictly zero, as illustrated in Fig.~1(d). Once these symmetry conditions are violated, pronounced SP emerges and reverses sign upon inverting molecular chirality, the hallmark of the chirality-induced spin selectivity (CISS) effect~\cite{Ray1999,Gohler2011,Naaman2019,Bloom2024,Chen2026,Wang2024}. The extended no-go theorem's predictions are thus systematically tested in CISS-related transport through chiral cyclodextrin assemblies.

The extended no-go theorem is verified in these chiral cyclodextrin devices, validating our theoretical approach and core mechanism for CISS in real chiral molecular systems. Since its discovery, the CISS mechanism has still remained debate~\cite{Evers2022,Nuomin2026}, and its study in realistic chiral supramolecular systems leaves a recognized theoretical gap. This work develops both a symmetry-protection theory for nonmagnetic open systems and a theoretical framework for CISS in realistic supramolecular systems, offering a new route to reveal its core mechanism.

\begin{figure}[t]
\includegraphics[width=0.99\columnwidth]{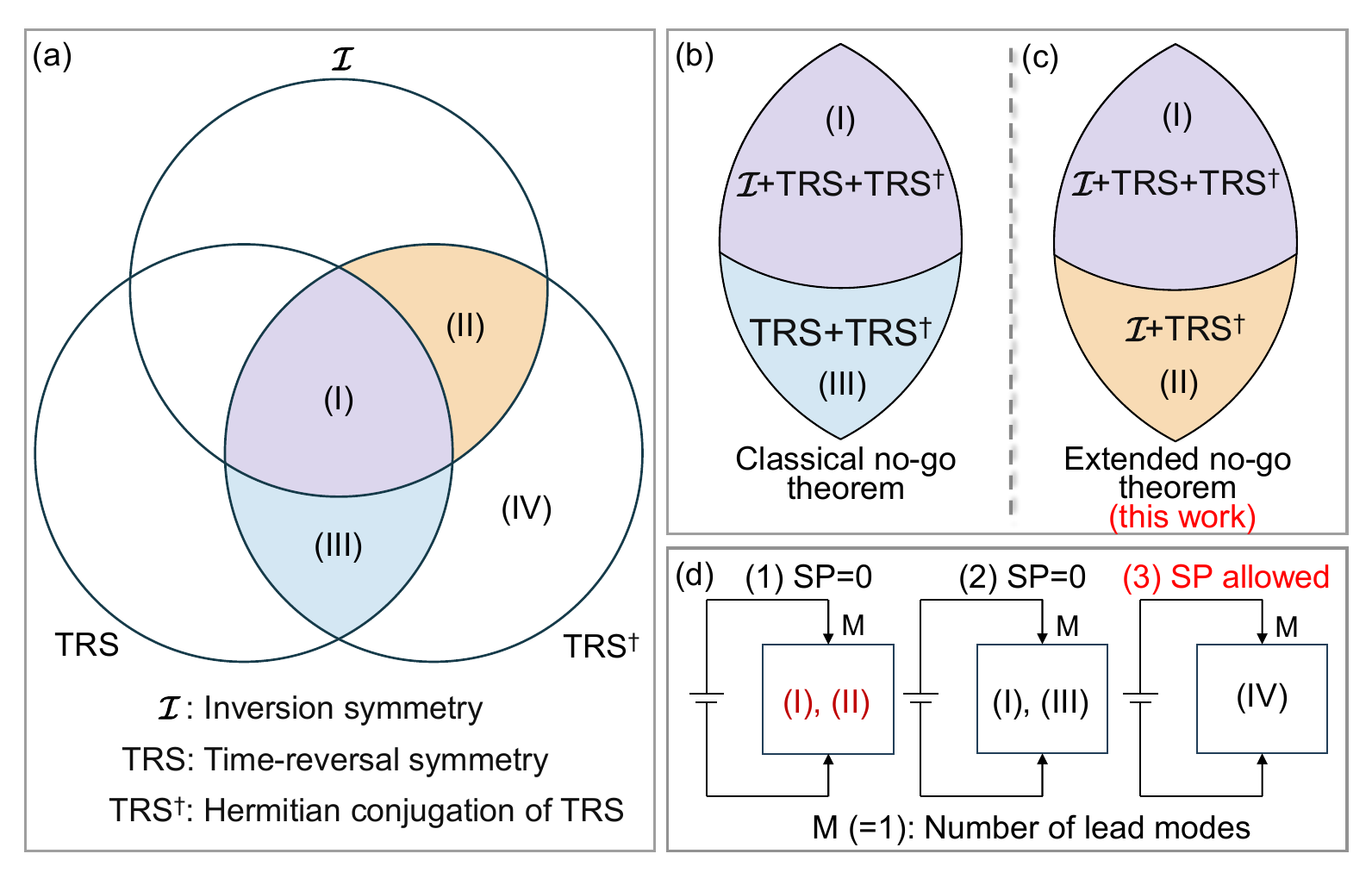}
\caption{(a) Relation among central inversion symmetry \(\mathcal{I}\), TRS, and symplectic reciprocity \(\mathrm{TRS}^{\dagger}\). Shaded regions mark where the two single-channel no-go theorems enforce \(\mathrm{SP}=0\). (b),(c) Validity regions of the classical [(I) and (III)] and extended [(I) and (II)] no-go theorems. (d) Three single-channel (\(M=1\), spin excluded) TRS devices. Devices (1) in (I) and (II), and (2) in (I) and (III) are protected by the extended and classical theorems, respectively, yielding \(\mathrm{SP}=0\) under unpolarized injection; device (3) in region (IV) allows nonzero \(\mathrm{SP}\).}
\label{fig1}
\end{figure}

{\color{blue}\emph{$\mathrm{TRS}$ and $\mathrm{TRS}^{\dagger}$ in open scattering.}}
For stationary scattering of spin-\(1/2\) particles, we use spin-independent ports normalized to unit flux and pair the incoming and outgoing channels according to their time-reversal partners~\cite{Brusheim2008,Jacquod2012,Beenakker1997}. The time-reversal operator is written as $\Theta=(I\otimes i\sigma_y)K\equiv JK$, where \(K\) denotes complex conjugation; \(I\) acts on the spin-independent transport channels; and \(J=i\sigma_y\) acts in spin space. The TRS constraint on the scattering matrix reads
\[
JS^*J^{-1}=S^{-1}. \tag{1}
\]
When the full Hamiltonian is Hermitian and all open scattering channels are included, \(S\) is unitary. Using \(S^\dagger=S^{-1}\) and \(J^\dagger=J^{-1}\), Eq.~(1) is equivalent to
\[
JS^TJ^{-1}=S. \tag{2}
\]
Defining the symplectic transpose \(S^R\equiv JS^TJ^{-1}\), we see that Eq.~(2) becomes \(S^R=S\), referred to as symplectic self-duality or symplectic reciprocity and denoted by $\mathrm{TRS}^{\dagger}$~\cite{Dyson1962,Mello1991,Kawabata2019}. Equations~(1) and~(2) are equivalent when \(S\) is unitary and correspond to different symmetries when \(S\) is nonunitary; conversely, the simultaneous validity of both equations under the same \(J\) also implies that \(S\) is unitary. Therefore, unitarity together with TRS is equivalent to TRS together with \(\mathrm{TRS}^{\dagger}\), which is precisely the condition of the classical two-terminal single-channel no-go theorem.

We partition the open channels into observed and unobserved sets, \(P\) and \(X\), each containing complete time-reversed pairs. The incoming and outgoing vectors are $a=(a_P,a_X)^T$ and $b=(b_P,b_X)^T$, and the total matrices have the block form
\[
S_{\mathrm{tot}}=\begin{pmatrix}s&U\\V_X&S_X\end{pmatrix},\qquad
J_{\mathrm{tot}}=\begin{pmatrix}J_P&0\\0&J_X\end{pmatrix}. \tag{3}
\]
Here \(s=S_{PP}\) is the scattering subblock at the observed terminals; \(V_X\) describes leakage from \(P\) to \(X\); \(U\) describes scattering from \(X\) to \(P\); and \(J_P\) and \(J_X\) represent the unitary parts of the time-reversal operators in the respective subspaces. Setting \(a_X=0\) gives \(b_P=sa_P\) and \(b_X=V_Xa_P\). The \(PP\) block of the complete symplectic self-duality relation yields $s=J_Ps^TJ_P^{-1}$, so \(s\) inherits \(\mathrm{TRS}^{\dagger}\).

However, \(s\) generally does not inherit TRS. Complete unitarity gives $s^\dagger s+V_X^\dagger V_X=I_P$, where \(V_X^\dagger V_X\) is the fraction of incident flux from the observed terminals that leaks into \(X\); for \(V_X\neq0\), \(s\) is no longer unitary. Taking the \(PP\) block of the full TRS relation and using the \(PX\) block of symplectic self-duality, \(U=J_PV_X^TJ_X^{-1}\), one obtains
\[
sJ_Ps^*=J_P\left[I_P-(V_X^\dagger V_X)^T\right]. \tag{4}
\]
As long as \(V_X\neq0\), we have \(sJ_Ps^*\neq J_P\), so the TRS in Eq.~(1) is broken. Physically, the outgoing amplitude flowing to \(X\) in the original process becomes an incoming amplitude from \(X\) after time reversal, which is incompatible with the boundary condition \(a_X=0\). Consequently, the reduced \(s\) retains only \(\mathrm{TRS}^{\dagger}\) and loses TRS.

After eliminating the unobserved degrees of freedom, environmental effects enter a non-Hermitian effective Hamiltonian~\cite{Ashida2020,YiYang2020,JinSong2021,YangWang2026}; when the port couplings are compatible with the relevant symmetry operations, the TRS and \(\mathrm{TRS}^{\dagger}\) of the Hamiltonian lead respectively to the two types of scattering constraints in Eqs.~(1) and~(2)~\cite{KawabataRyu2021}. Accordingly, the classical no-go theorem relies on TRS and unitarity, equivalently on TRS and \(\mathrm{TRS}^{\dagger}\); in contrast, this work employs TRS\(^\dagger\) together with the central inversion \(\mathcal{I}\) that exchanges the two terminals as another set of sufficient protection conditions. The two sets of conditions intersect but neither contains the other. In what follows, we prove the conclusions of this work directly under this scattering constraint.

\begin{figure*}[t]
\includegraphics[width=1.95\columnwidth]{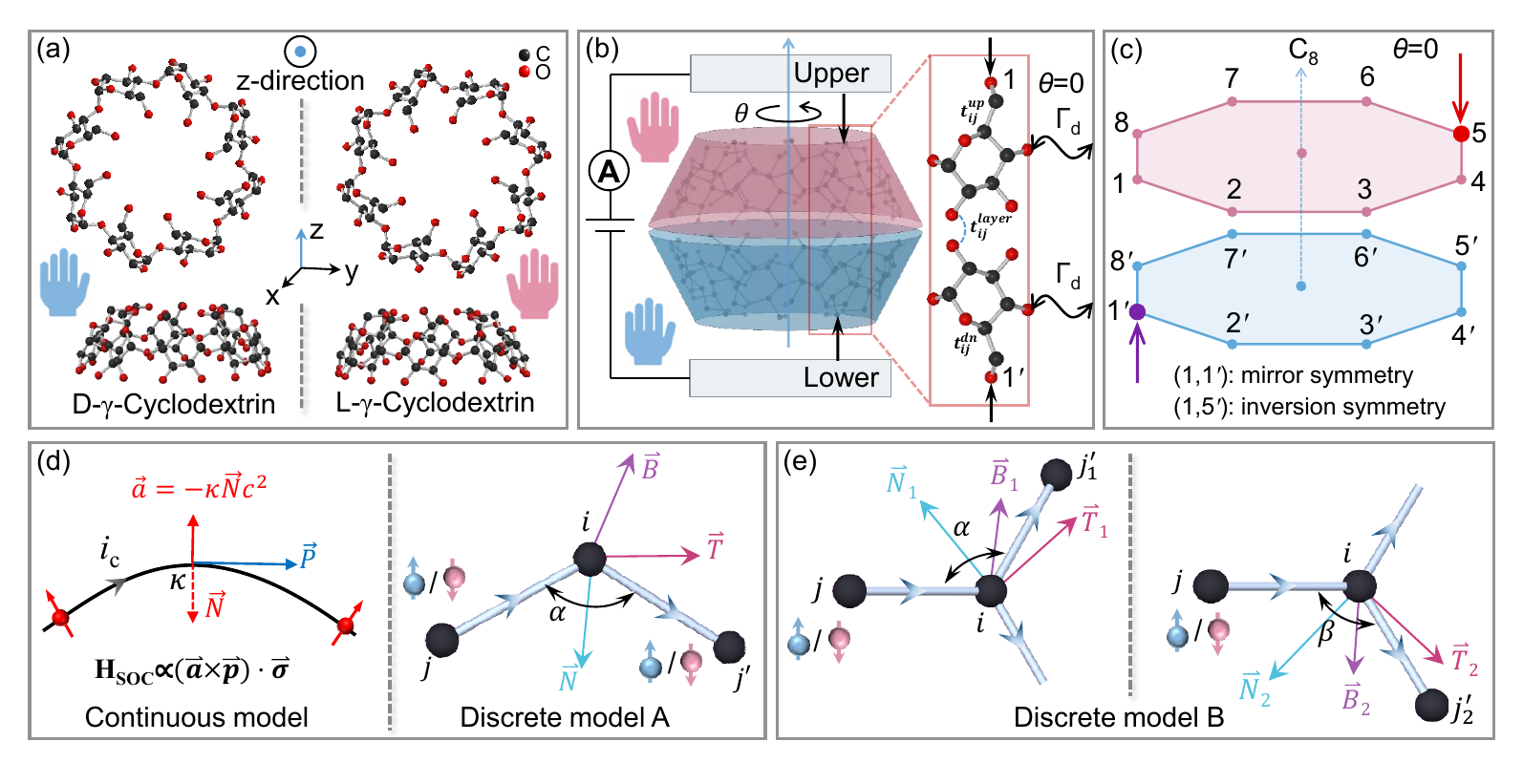}
\caption{(a) D- and L-enantiomers of \(\gamma\)-cyclodextrin (H atoms omitted); top and side views. Black and red dots denote C and O atoms. (b) (b) Coaxial D–L dimer with the large rims facing each other, coupled to upper (U) and lower (L) electrodes. The upper molecule and its electrode rotate as a whole by an angle \(\theta\) relative to the lower molecule; \(\theta=0\) corresponds to mirror-symmetric contacts at sites \((1,1')\). \(\Gamma_d\) denotes environmental broadening. (c) Electrode contact sites \(m\) and \(n'\). (d) Continuous geometric SOC along a curved path and its discretized form; \(\vec{T}\), \(\vec{N}\), and \(\vec{B}\) are the Frenet--Serret tangent, normal, and binormal vectors. (e) Discrete geometric SOC at a branching path.}
\label{fig2}
\end{figure*}

{\color{blue}\emph{Extended No-Go Theorem.}}
In the two-terminal scattering model, we label the upper and lower terminals by \(U\) and \(L\), respectively, and write the scattering matrix as
\[
S=\begin{pmatrix}r_L&t_{LU}\\t_{UL}&r_U\end{pmatrix}. \tag{5}
\]
No unitarity of \(S\) is assumed here. Since the central inversion symmetry imposed below exchanges the two terminals, the numbers of channel modes at the two terminals must be equal; after spin is included, each block of \(S\) is a \(2M\times2M\) matrix. The TRS\(^\dagger\) condition in Eq.~(2) gives, for the off-diagonal blocks,
\[
t_{LU}=J_Mt_{UL}^TJ_M^{-1}\equiv t_{UL}^R, \tag{6}
\]
where \(J_M=I_M\otimes i\sigma_y\). Equation~(6) relates the transmission matrices for opposite transport directions by a symplectic transpose. We now assume that the full device is invariant under central inversion, which exchanges the two terminals. Spin is an axial vector and remains invariant under spatial inversion; hence spatial inversion acts as the identity in spin space. In the time-reversal-compatible basis gauge, we choose the mode bases of the upper and lower terminals so that they are mapped onto each other under inversion. The spatial inversion operator then reads
\[
P=\begin{pmatrix}0&I_{2M}\\I_{2M}&0\end{pmatrix}. \tag{7}
\]
Central inversion symmetry requires $PSP^{-1}=S$. Multiplying blockwise yields
\[
\begin{pmatrix}r_U&t_{UL}\\t_{LU}&r_L\end{pmatrix}
=
\begin{pmatrix}r_L&t_{LU}\\t_{UL}&r_U\end{pmatrix}. \tag{8}
\]
Therefore \(r_L=r_U\) and \(t_{LU}=t_{UL}\equiv t\). Combining this with Eq.~(6) gives \(t=t^R\), i.e., the transmission amplitude matrix for a given direction also possesses symplectic self-duality.

When \(M=1\), \(t\) is a \(2\times2\) matrix, and its symplectic transpose is
\[
t^R=\begin{pmatrix}t_{22}&-t_{12}\\-t_{21}&t_{11}\end{pmatrix}. \tag{9}
\]
Elementwise, \(t=t^R\) requires \(t_{11}=t_{22}\) and \(t_{12}=t_{21}=0\), so \(t\propto I\). This completes the proof of the no-go theorem in this work. Let \(\rho_{\mathrm{in}}\) be the spin density matrix of the incident electron; the unnormalized transmitted state is \(\bar\rho_{\mathrm{out}}=t\rho_{\mathrm{in}}t^\dagger\). When the transmission probability is nonzero, normalization gives \(\rho_{\mathrm{out}}=\rho_{\mathrm{in}}\), i.e., the transmission process preserves the incident spin state; if the incident state is unpolarized, the outgoing state is also unpolarized. This result is stronger than the usual statement \(t^\dagger t\propto I\) of the classical no-go theorem~\cite{Jiang2013,Bulgakov2002,Kiselev2003}.

\begin{figure*}[t]
\includegraphics[width=1.985\columnwidth]{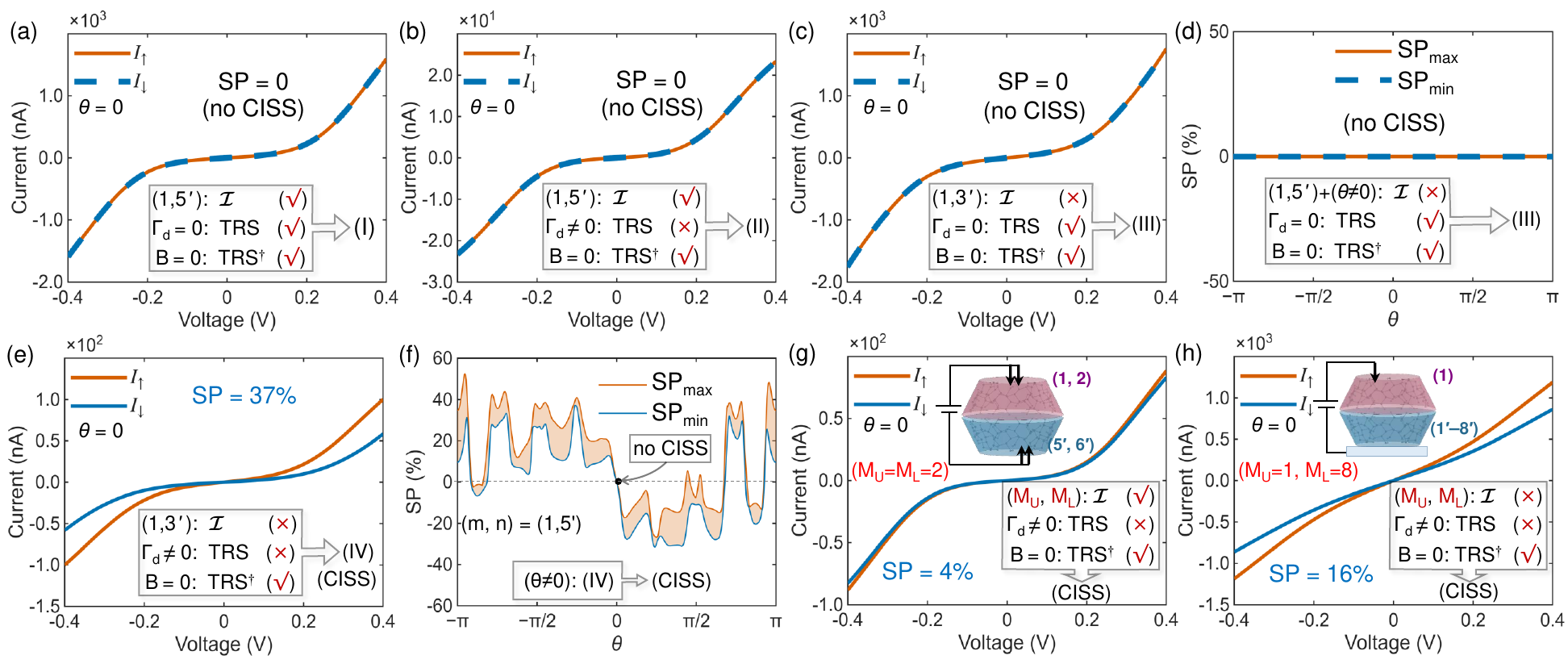}
\caption{Spin polarization (SP) and spin-dependent currents in cyclodextrin-assembly devices. (a)-(c) Spin-dependent currents versus bias voltage at \(\theta=0\) for contact configurations (I)-(III) in the single-orbital mode. Panels (a) and (b) correspond to the terminal-connected model \((1,5^{\prime})\) with \(\Gamma_d=0\), and panel (c) to model \((1,3^{\prime})\) with \(\Gamma_d=0\). (d) SP versus \(\theta\) in the two-terminal device with model \((1,5^{\prime})\) and \(\Gamma_d=0\). (e) Current versus bias voltage for model \((1,3^{\prime})\) at \(\theta=0\). (f) SP versus \(\theta\) for model \((1,5^{\prime})\) with \(\Gamma_d=0.015\,\mathrm{eV}\), where the upper molecule and electrode are rotated. (g) Current versus bias voltage in the two-terminal device at \(\theta=0\), with \(M_u=(1,2)\), \(M_L=(5^{\prime},6^{\prime})\), and \(\Gamma_d=0.015\,\mathrm{eV}\). (h) Geometry for CISS measurement in a real device: a top tip contacts site 1 of the upper molecule, and all eight lower sites couple to the substrate with \(\Gamma_d=0.015\,\mathrm{eV}\) and \(\theta=0\).}
\label{fig3}
\end{figure*}

When the number of modes per terminal increases to \(M=2\), \(t\) becomes a \(4\times4\) matrix, and \(t=t^R\) is no longer sufficient to exclude spin-selective transmission~\cite{Utsumi2020,Perroni2007,Santos2020,Korytar2024,DOnofrio2025,Guo2016}. To characterize the continuous response after the single-channel protection is lifted, we employ a two-channel contact-coupling model in which the second mode is decoupled at \(q=0\) and gradually participates in transport for \(q>0\), while the total broadening at each terminal remains fixed. The two broadening eigenvalues are \(g_+=2\gamma/(1+q)\) and \(g_-=2\gamma q/(1+q)\), with \(q=g_-/g_+\in[0,1]\); central inversion symmetry ensures that the corresponding modes at the upper and lower terminals are paired with equal broadening. For the contact family defined above, the Fisher–Lee relation~\cite{FisherLee1981} yields \(\Delta T=c_z(E,\lambda)q+O(q^2)\) for fixed \(\lambda\) and \(\Delta T=d_z(E,q)\lambda^2+O(\lambda^3)\) for fixed \(q\). Explicit definitions of the contact-coupling and broadening matrices, the coefficients \(c_z\) and \(d_z\), together with the detailed derivation, are provided in the End Matter; the main text retains only the above conclusions, and we examine the validity of this small-parameter expansion using devices constructed from realistic materials.

%When the number of modes per terminal increases to \(M=2\), \(t\) becomes a \(4\times4\) matrix, and \(t=t^R\) is no longer sufficient to exclude spin-selective transmission. To characterize the continuous response after the single-channel protection is lifted, we employ a two-channel contact-coupling model in which the second mode is decoupled from the central region at \(q=0\) and gradually participates in transport for \(q>0\), while the total broadening at each terminal is kept fixed. The two broadening eigenvalues are taken as \(g_+=2\gamma/(1+q)\) and \(g_-=2\gamma q/(1+q)\), with \(q=g_-/g_+\in[0,1]\); central inversion symmetry ensures that the corresponding modes at the upper and lower terminals are paired with equal broadening. Under the wide-band approximation, purely imaginary contact self-energies, and the Fisher-Lee relation, one obtains \(\Delta T=c_z(E,\lambda)q+O(q^2)\) for fixed \(\lambda\) and \(\Delta T=d_z(E,q)\lambda^2+O(\lambda^3)\) for fixed \(q\). The explicit definitions of the contact-coupling and broadening matrices, the coefficients \(c_z\) and \(d_z\), and the auxiliary functions \(f_j\) and \(F_j\), together with the detailed derivation, are provided in the Appendix; only the above conclusions are retained in the main text, where we also state that the range of validity of this small-parameter expansion will be examined using devices constructed from realistic materials. 

{\color{blue}\emph{Chiral cyclodextrin-based transport devices.}}
Next, we use two chiral enantiomers of \(\gamma\)-cyclodextrin to construct a two-terminal device and thereby test the extended no-go theorem and its corollaries. As shown in Fig.~2(a) (all atomic sites adopt their actual coordinates~\cite{Pereva2024}), the \(\gamma\)-cyclodextrin molecule has a hollow truncated-cone shape~\cite{Harata1987,CyclodextrinReview2022} and possesses \(C_8\) rotational symmetry; it is a point-symmetric chiral molecule composed of eight repeat units. The natural molecule exhibits D chirality, and its L enantiomer has been successfully synthesized experimentally~\cite{Wu2024MirrorCyclodextrins}. We stack D- and L-cyclodextrin coaxially with their large ends facing each other to form a chiral assembly; such superstructures have precedents~\cite{KazemRostami2022,Christoforides2018}. Electrodes are connected to the upper and lower ends of this assembly (Fig.~2(b)) to construct a double-layer molecular heterojunction transport device. When the relative orientation of the D and L molecules is appropriate, the coaxial dimer exhibits mirror symmetry (\(\sigma_h\)). Since the dimer also retains twofold rotational symmetry about this axis, the spatial operation \(C_{2z}\sigma_h\) is equivalent to central inversion. By choosing the electrode contact positions to preserve this symmetry and including environment coupling, we construct a complete device with central inversion symmetry (Fig.~2(c)), which can be used to test the extended no-go theorem and its derivations.

%Next, we use two chiral enantiomers of \(\gamma\)-cyclodextrin to construct a two-terminal device and thereby test the extended no-go theorem and its corollaries. As shown in Fig.~2(a) (all atomic sites adopt their actual coordinates), the \(\gamma\)-cyclodextrin molecule has a hollow truncated-cone shape and possesses \(C_8\) rotational symmetry; it is a point-symmetric chiral molecule composed of eight repeat units. The natural molecule exhibits D chirality, and its L enantiomer has been successfully synthesized experimentally. We stack D- and L-cyclodextrin coaxially with their large ends facing each other to form a chiral assembly; superstructures stacked in this large-end-facing manner have precedents. Electrodes are connected to the upper and lower ends of this assembly (Fig.~2(b)) to construct a double-layer molecular heterojunction transport device. When the relative orientation of the D and L molecules is appropriate, the coaxial dimer exhibits mirror symmetry, with the corresponding mirror reflection operation \(\sigma_h\). Since the dimer also retains twofold rotational symmetry about this axis, the spatial operation \(C_{2z}\sigma_h\) is equivalent to central inversion. By choosing the electrode contact positions to preserve this symmetry and including environment coupling, we construct a complete device with central inversion symmetry (Fig.~2(c)), which can be used to test the extended no-go theorem and its corollaries.

Cyclodextrins comprise light atoms (C, O, H), whose weak intrinsic SOC cannot produce observable SP. We find that electrons traversing these chiral molecules move along curved or folded paths. Along a curved trajectory, the continuously changing direction of motion produces a relativistic effect in the non-inertial frame, constraining the spin orientation and generating local SP, as explained in Fig.~2(d). This manifests as geometric SOC, governed by the local curvature~\cite{Shitade2020,Yu2024}. In the Frenet-Serret frame, the local inertial acceleration at site \(i\) is $\mathbf{a}_i$=$-\kappa_i c^2\mathbf{N}_i$, with \(\kappa_i\) the local curvature and \(\mathbf{N}_i\) the normal vector toward the curvature center. The geometric SOC Hamiltonian is $\mathcal{H}_{\mathrm{geo}}$=$\sum_i\frac{\hbar}{4mc^2}(\mathbf{a}_i\times\mathbf{p}_i)\cdot\boldsymbol{\sigma}$, where \(\mathbf{p}_i\), \(\boldsymbol{\sigma}\), \(m\), and \(c\) denote the momentum operator, Pauli matrices, electron mass, and light speed, respectively~\cite{HehlNi1990,Yu2024}. For the two specific electron motion modes within cyclodextrin, as described in Figs.~2(d) and 2(e), we second-quantize and discretize the geometric SOC, establishing a site-resolved geometric SOC Hamiltonian \(\mathcal{H}_{\mathrm{SOC}}\). Its derivation and explicit form are provided in the End Matter.

%Since cyclodextrin molecules consist only of light atoms such as C, O, and H, their weak intrinsic atomic spin-orbit coupling (SOC) is insufficient to produce observable spin polarization. We find that when electrons transport through such chiral molecules, they move along curved or folded paths. As electrons move along a curved trajectory, their direction of motion changes continuously, producing a relativistic effect in the non-inertial frame; this effect constrains the spin orientation and generates local spin polarization (SP) [Fig.~2(d)]. This effect manifests as a geometric SOC, determined by the local curvature of the trajectory. In the Frenet-Serret frame, the local inertial acceleration at site \(i\) is \(\mathbf{a}_i=-\kappa_i c^2\mathbf{N}_i\), where \(\kappa_i\) is the local curvature and \(\mathbf{N}_i\) is the normal vector pointing toward the center of curvature. The geometric SOC Hamiltonian is \(\mathcal{H}_{\mathrm{geo}}=\sum_i\frac{\hbar}{4mc^2}(\mathbf{a}_i\times\mathbf{p}_i)\cdot\boldsymbol{\sigma}\), where \(\mathbf{p}_i\) is the momentum operator, \(\boldsymbol{\sigma}\) is the vector of Pauli matrices, \(m\) is the electron mass, and \(c\) is the speed of light. Furthermore, based on the two specific modes of electron motion within the cyclodextrin molecule [Figs.~2(d) and 2(e)], we second-quantize and discretize the geometric SOC to establish a site-resolved geometric SOC Hamiltonian \(\mathcal{H}_{\mathrm{SOC}}\). The derivation and explicit form are provided in the End Matter.

Using the site-resolved \(\mathcal{H}_{\mathrm{geo}}\), we construct a device model for this molecular junction [Fig.~2(a)]. Within the tight-binding framework, the total Hamiltonian is $\mathcal{H}$=$\mathcal{H}_{\mathrm{mol}}+\mathcal{H}_{\mathrm{geo}}+\mathcal{H}_{\mathrm{el}}+\mathcal{H}_{\mathrm{d}}$, where \(\mathcal{H}_{\mathrm{mol}}\) describes the on-site energies and electron hopping within the molecular assembly, \(\mathcal{H}_{\mathrm{el}}\) describes the coupling between the molecule and the upper and lower electrodes, and \(\mathcal{H}_{\mathrm{d}}\) describes the coupling between the molecule and environmental degrees of freedom. The explicit forms of these Hamiltonians, the model parameters, the electrode and environment self-energies are given in the End Matter. Finally, using the nonequilibrium Green's function method~\cite{Khomyakov2005,Beenakker1997,Guo2012}, we calculate the spin-resolved transmission coefficient \(T(E)\) for unpolarized electrons incident from the lower (L) electrode and outgoing from the upper (U) electrode, along with the transmission difference \(\Delta T(E)\), and then obtain the spin-dependent transport currents \(I_{\uparrow}\) and \(I_{\downarrow}\). The SP of the current contributed by the two observed terminals is defined as $\mathrm{SP}=\frac{I_{\uparrow}-I_{\downarrow}}{I_{\uparrow}+I_{\downarrow}}$. The calculation methods for charge currents are detailed in the End Matter.

%After constructing the site-resolved \(\mathcal{H}_{\mathrm{geo}}\), we establish a theoretical model for the device based on this molecular junction [Fig.~2(a)]. Within the tight-binding framework, the total Hamiltonian of the system is written as $\mathcal{H}=\mathcal{H}_{\mathrm{mol}}+\mathcal{H}_{\mathrm{geo}}+\mathcal{H}_{\mathrm{el}}+\mathcal{H}_{\mathrm{d}}$, where \(\mathcal{H}_{\mathrm{mol}}\) describes the on-site energies and electron hopping within the molecular assembly, \(\mathcal{H}_{\mathrm{el}}\) describes the coupling between the molecule and the upper and lower electrodes, and \(\mathcal{H}_{\mathrm{d}}\) describes the coupling between the molecule and environmental degrees of freedom. The explicit forms of these Hamiltonians, the model parameters, the electrode and environment self-energies, and the Green's functions are given in the End Matter. Finally, using the nonequilibrium Green's function method, we calculate the spin-resolved transmission coefficient \(T(E)\) for unpolarized electrons incident from the lower (L) electrode and outgoing from the upper (U) electrode, along with the transmission difference \(\Delta T(E)\), and then obtain the spin-dependent transport currents \(I_{\uparrow}\) and \(I_{\downarrow}\). The spin polarization (SP) of the current contributed by the two observed terminals is defined as $\mathrm{SP}=\frac{I_{\uparrow}-I_{\downarrow}}{I_{\uparrow}+I_{\downarrow}}$. The calculation methods for the above Hamiltonians and charge currents are detailed in the End Matter. 

To validate the use of chiral cyclodextrin molecules for studying the CISS effect, we construct a single-molecule tunneling device (Sec.~II of the Supplemental Material)~\cite{SupplementalMaterial}. Calculations show that a directional current through individual D- or L-cyclodextrin molecules produces significant SP, which reverses sign upon chirality inversion. This confirms that our Hamiltonian model and computational method effectively capture the CISS effect in chiral cyclodextrin molecules.

%To verify whether chiral cyclodextrin molecules can be used to study the CISS effect, we construct a tunneling device with a single cyclodextrin molecule in Sec.~II of the Supplemental Material. The calculated results show that a directional current through a single D- or L-type cyclodextrin molecule indeed produces a significant spin polarization, and the polarization direction reverses when the chirality is reversed. This demonstrates that our Hamiltonian model and computational method can very effectively study the CISS effect in chiral cyclodextrin molecules.

{\color{blue}\emph{To prove the theorem in realistic materials.}}
We examine the extended no-go theorem and its key corollaries in cyclodextrin chiral molecular assembly devices by tuning the single-channel electrode connection configuration \((m,n')\), the leakage broadening \(\Gamma_d\), and the assembly angle \(\theta\). The cyclodextrin molecules are nonmagnetic and \(\mathbf B=0\), so generalized time-reversal symmetry (TRS\(^+\)) holds; the effective symmetry relevant to transport is determined jointly by these parameters. A single-channel connection is denoted by \((m,n')\) [Figs.~2(b) and 2(c)]. For \((m,n')=(1,5')\), the device possesses inversion symmetry \(I\). When \(\Gamma_d=0\), the effective TRS condition is preserved, corresponding to region (I) in Fig.~1(a); a finite \(\Gamma_d\) (e.g., 0.05~eV) breaks this effective TRS condition and corresponds to region (II). According to the theorem, the spin polarization \(\mathrm{SP}\) should vanish strictly in both regions. Numerical calculations show that \(I_\uparrow\) and \(I_\downarrow\) coincide throughout the bias range shown, consistent with the predicted zero spin polarization. For \((1,3')\), \(\mathcal{I}\) is broken while \(\Gamma_d=0\) still preserves the effective TRS condition; the calculated \(\mathrm{SP}\) remains zero [Fig.~3(c)], corresponding to region (III). Similarly, for \((1,5')\), \(\Gamma_d=0\), and \(\theta\neq0\), \(I\) is broken and the system remains in region (III), with \(\mathrm{SP}\) still strictly zero [Fig.~3(d)]. These results correspond to regions (I), (II), and (III) in Fig.~1 and agree with the predictions of the extended no-go theorem.

Once the device parameters depart from the above three regions, the spin polarization of the charge current is no longer constrained by the no-go theorem, and the chiral-induced spin selectivity (CISS) effect can emerge. A typical case is breaking inversion symmetry while keeping \(\Gamma_d\) finite: for example, \((1,3')\) with \(\Gamma_d=0.05\)~eV places the system in region (IV) of Fig.~1, where \(\mathrm{SP}=0\) is no longer guaranteed. Transport calculations yield \(\mathrm{SP}=37.0\%\), and the polarization reverses sign for the opposite chirality [See SM], directly confirming the CISS effect. Similarly, for the connection \((1,5')\), setting \(\theta\) to a nonzero value also breaks the global inversion symmetry and enables the CISS effect [Fig.~3(f); results for the opposite chirality are shown in Fig.~S5]. Furthermore, multichannel connections are no longer protected by the extended no-go theorem against \(\mathrm{SP}=0\) and are thus more favorable for the CISS effect. Changing the single-channel connection in Fig.~3(b) to upper-electrode connections at sites 1 and 2 and lower-electrode connections at sites \(5'\) and \(6'\) yields a nonzero \(\mathrm{SP}=4\%\), with the polarization reversing sign for the opposite chirality (Fig.~S6). If the upper molecule connects to the electrode through only one atom while all lowest sites of the lower molecule contact the metal substrate, the resulting asymmetric contact configuration is the most common configuration in CISS experiments~\cite{Kettner2015,Firouzeh2024,Albro2025}. This configuration is not subject to the \(\mathrm{SP}=0\) constraint, and the CISS effect occurs regardless of the upper-electrode connection site, as also confirmed by our transport calculations [Fig.~3(h)].

\begin{figure}[t]
\includegraphics[width=0.99\columnwidth]{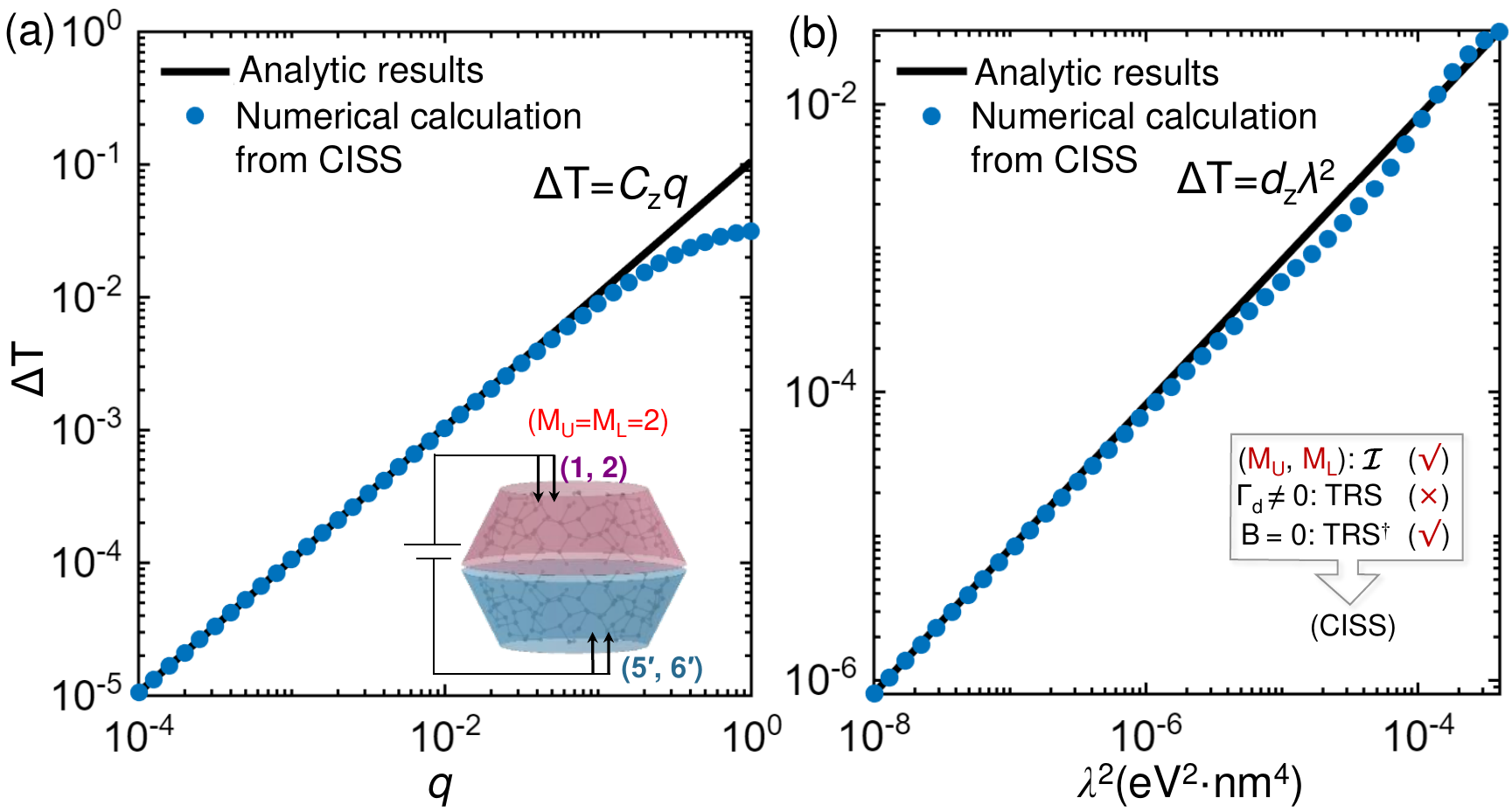}
\caption{Using the contact and leakage settings of Fig.~3(e) with inversion-paired contact modes, at \(E=-1.58\,\mathrm{eV}\): (a) \(\Delta T\) versus the broadening ratio \(q\) for fixed \(\lambda=0.02\,\mathrm{eV}\cdot\mathrm{nm}^2\); (b) \(\Delta T\) versus \(\lambda^2\) for fixed \(q=1\). Blue dots: CISS transport calculations for the cyclodextrin assembly; black lines: leading-order analytical results of the extended no-go theorem.}
\label{fig4}
\end{figure}

Finally, we test two key corollaries of the extended no-go theorem: for \(M=2\), \(\Delta T\) should scale linearly with \(q\) and \(\lambda^2\), respectively [Figs.~4(a) and 4(b)]. The device adopts a two-channel configuration with the upper electrode connected to molecular sites 1 and 2 and the lower electrode to sites \(5'\) and \(6'\), and \(\Gamma_d\neq0\) (e.g., 0.05~eV). To retain, at \(q=0\), a single contact mode that still covers both contact atoms, we linearly combine the couplings of the two electrodes: \(|+\rangle=(|\alpha_1\rangle+|\alpha_2\rangle)/\sqrt{2}\) and \(|-\rangle=(|\alpha_1\rangle-|\alpha_2\rangle)/\sqrt{2}\), where \(\alpha=U,L\), and \(|\alpha_{1,2}\rangle\) denote the orbitals of the atoms contacted by the upper and lower electrodes. From the transmission spectrum we select a representative energy \(E=-1.58\)~eV (Sec.~3 of the Supplemental Material). We calculate \(\Delta T\) for this actual cyclodextrin assembly device and present its dependence on \(q\) (with \(\lambda^2\) fixed) and on \(\lambda^2\) (with \(q\) fixed) [Figs.~4(a) and 4(b)]. Over a wide parameter range, the numerical results agree closely with the theoretical analysis. In summary, transport calculations for actual devices under various symmetry conditions agree with the extended no-go theorem and its corollaries, validating the theorem and supporting the correctness of identifying geometric spin-orbit coupling (SOC) as the core mechanism of the CISS effect, as well as the correctness of our theoretical framework for studying the CISS effect in real chiral material systems.

In summary, we have established an extended no-go theorem for spin-selective transport in nonmagnetic open scattering systems and verified it in chiral cyclodextrin assemblies. By separating full-system symmetry from constraints on the reduced observed scattering matrix, we showed that TRS and symplectic reciprocity TRS\(^{\dagger}\) coincide only for unitary scattering, and that leakage into unobserved channels breaks TRS while preserving TRS\(^{\dagger}\). When the reduced scattering satisfies TRS and the full device preserves central inversion exchanging the terminals, the single-channel transmission matrix must be a spin scalar, enforcing strictly zero spin polarization. Transport calculations in \(\gamma\)-cyclodextrin enantiomer devices confirm this protection and show that broken symmetry yields chirality-reversed spin polarization, with two-channel scaling in agreement with analytical predictions. These results provide a symmetry-based framework for spin transport in open nonunitary systems and a predictive tool for the chirality-induced spin selectivity effect.

{\color{blue}\emph{Acknowledgments.}} This work is supported by the authors' personal resource.

\newpage
\section*{End Matter}

\subsection*{Continuous response for \(M=2\)}

When the number of modes per terminal increases to \(M=2\), \(t\) becomes a \(4\times4\) matrix, and Eq.~(9) is no longer sufficient to rule out spin-selective transmission. To study the continuous response after the single-channel protection is lifted, we keep the contact-mode basis fixed in a two-channel representation, gradually increase the coupling of the second mode to the central region, and keep the total broadening at each terminal fixed. The second mode is decoupled from the central region at \(q=0\), and both modes participate in transport for \(q>0\). Denoting the always-open mode by \(+\) and the mode closed at \(q=0\) by \(-\), we take the broadening eigenvalues of the two modes coupled to the central scattering region as
\begin{equation}
g_+=\frac{2\gamma}{1+q},\qquad
g_-=\frac{2\gamma q}{1+q}, \tag{10}
\end{equation}
where \(2\gamma\) is the total broadening, \(g_++g_-=2\gamma\), and \(q=g_-/g_+\in[0,1]\) is the ratio of the two broadening eigenvalues. By central inversion symmetry, the two modes at the upper and lower terminals are paired with equal broadening eigenvalues. In the basis introduced above, let \(u_{\alpha,\pm}\) be the orthonormal orbital vectors coupling the modes at terminal \(\alpha=U,L\) to the central region, and define \(U_{\alpha,\pm}=u_{\alpha,\pm}\otimes I_2\). Using the standard contact-coupling representation [123],
\begin{equation}
W_\alpha=\bigl[\sqrt{g_+}U_{\alpha,+}\ \sqrt{g_-}U_{\alpha,-}\bigr],\qquad \tag{11}
\end{equation}
\begin{equation}
\Gamma_\alpha=W_\alpha W_\alpha^\dagger
=g_+U_{\alpha,+}U_{\alpha,+}^\dagger
+g_-U_{\alpha,-}U_{\alpha,-}^\dagger, \tag{12}
\end{equation}
where \(W_\alpha\) is the coupling matrix between electrode \(\alpha\) and the device, and \(\Gamma_\alpha\) is the broadening matrix.

We split the Hamiltonian into a spin-independent part and an SOC part, \(H=H_0+\lambda V\), where \(\lambda\) is the SOC coupling strength and \(V\) is the SOC part, the only spin-dependent part of the system. In the wide-band approximation with energy-independent purely imaginary contact self-energies~\cite{Verzijl2013}, the retarded Green's function is
\begin{equation}
G^r(E,q,\lambda)=\bigl[EI-H-\Sigma_X^r-\Sigma_L^r-\Sigma_U^r\bigr]^{-1}, \tag{13}
\end{equation}
where \(\Sigma_X^r\) is the environment self-energy independent of \(q\) and \(\lambda\), and \(\Sigma_\alpha^r=-\frac{i}{2}\Gamma_\alpha\) are the contact self-energies at the two terminals. The device preserves TRS\(^\dagger\) and central inversion symmetry. We define the difference between the spin-up and spin-down transmission coefficients as
\begin{equation}
\Delta T=\mathrm{Tr}\bigl[(I_2\otimes\sigma_z)tt^\dagger\bigr]. \tag{14}
\end{equation}
Using the Fisher-Lee relation \(t=-iW_U^\dagger G^rW_L\), we obtain the following asymptotic results (see the Appendix for details). For fixed \(\lambda\) and small \(q\),
\begin{equation}
\Delta T=c_z(E,\lambda)q+O(q^2), \tag{15}
\end{equation}
where
\begin{equation}
c_z(E,\lambda)=-32\gamma^2\,\mathrm{Im}\bigl[f_x(E,0,\lambda)f_y(E,0,\lambda)^*\bigr], \tag{16}
\end{equation}
\begin{equation}
f_j(E,q,\lambda)=\frac12\mathrm{Tr}\bigl[\sigma_jU_{U+}^\dagger G^r(E,q,\lambda)U_{L-}\bigr],\quad j=x,y,z. \tag{17}
\end{equation}
For fixed \(q\) and small \(\lambda\),
\begin{equation}
\Delta T=d_z(E,q)\lambda^2+O(\lambda^3), \tag{18}
\end{equation}
where
\begin{equation}
d_z(E,q)=-\frac{32\gamma^2q}{(1+q)^2}\,\mathrm{Im}\bigl[F_x(E,q)F_y(E,q)^*\bigr], \tag{19}
\end{equation}
\begin{equation}
F_j(E,q)=\frac12\mathrm{Tr}\bigl[\sigma_jU_{U+}^\dagger G^r(E,q,0)VG^r(E,q,0)U_{L-}\bigr]. \tag{20}
\end{equation}
These results assume that the Green's function exists at the relevant energy and that the low-order expansion coefficients are nonzero. The range of validity of this small-parameter expansion is examined numerically in the main text using realistic device structures.

\subsection*{Discrete Geometric SOC}

In a real cyclodextrin molecule, local transport paths can be classified into two types (Fig.~2(d) right panel and Fig.~2(e)): a single-channel folded path passing through site \(i\) with bond angle \(\alpha\neq\pi\) (path a: \(j\to i\to j'\)), and a branched path where three bonds intersect at site \(i\) with angles \(\alpha\) and \(\beta\) (path b: \(j\to i\to j'_{1(2)}\)). For these local single-channel and branched paths, the tight-binding discretization gives
\[
\mathcal{H}_{\mathrm{geo}}
=i\frac{\lambda}{2}\sum_i\sum_{j\in(i)}f(n_i)
\sum_{j'\in(i)}^{j'\neq j}
c_i^\dagger
\frac{\kappa_{jij'}\boldsymbol{\sigma}_{jij'}\cdot\mathbf{B}_{jij'}}{l_{jij'}}
c_j+\mathrm{H.c.}, \tag{21}
\]
where \(\lambda=\hbar^2/(4m)\simeq0.02~\mathrm{eV}\,\mathrm{nm}^2\); \(f(n_i)=1/(n_i-1)\) normalizes the branch-path contributions at site \(i\), with \(n_i\ge2\) and \(n_i\) the corresponding coordination number; \(c_j^\dagger=(c_{j,\uparrow}^\dagger,c_{j,\downarrow}^\dagger)\) is the electron creation operator at site \(j\); \(\kappa_{jij'}\) and \(\mathbf{B}_{jij'}\) are the local curvature and binormal unit vector of path a, respectively; \(l_{jij'}=l_{ji}+l_{ij'}\); and \(j,j'\in(i)\) denote the two endpoints of a path passing through site \(i\). For the single-channel case, the effective curvature is defined as \(\kappa_{jij'}=1/R_{jij'}\), where \(R_{jij'}\) is the radius of the circumcircle passing through sites \(j\), \(i\), and \(j'\). All bond lengths and bond angles are taken from the actual molecular structure and optimized to a stable geometry by first-principles calculations (Sec.~I of the SM, including the derivation of the above equations).

\subsection*{Model Hamiltonian of the Devices}

After constructing the site-resolved \(\mathcal{H}_{\mathrm{geo}}\), we can establish a theoretical model for the device based on this molecular junction (Fig.~2(a)). Within the tight-binding framework, the total Hamiltonian is $\mathcal{H}=\mathcal{H}_{\mathrm{mol}}+\mathcal{H}_{\mathrm{geo}}+\mathcal{H}_{\mathrm{el}}+\mathcal{H}_{\mathrm{d}}$, where \(\mathcal{H}_{\mathrm{mol}}\) describes the on-site energies and electron hopping within the molecular assembly, \(\mathcal{H}_{\mathrm{el}}\) describes the coupling between the molecule and the upper and lower electrodes, and \(\mathcal{H}_{\mathrm{d}}\) describes the coupling between the molecule and environmental degrees of freedom.  The on-site energies and electron hopping within the molecular assembly are described by
\begin{equation}
\mathcal{H}_{\mathrm{mol}}
=\sum_i\epsilon_i c_i^\dagger c_i
+\sum_{\langle ij\rangle}t_{ij}^{\mathrm{layer}}c_i^\dagger c_j
+\sum_{\langle ij\rangle\in\mathrm{up}}t_{ij}^{\mathrm{up}}c_i^\dagger c_j\\
+\sum_{\langle ij\rangle\in\mathrm{dn}}t_{ij}^{\mathrm{dn}}c_i^\dagger c_j
+\mathrm{H.c.} \tag{22}
\end{equation}
Here \(\epsilon_i\) are the on-site energies set by \(\epsilon_{\mathrm{C}}=0\) and \(\epsilon_{\mathrm{O}}=-2~\mathrm{eV}\); the second to fourth terms describe the interlayer hopping and the nearest-neighbor hopping within the upper and lower molecules, respectively. \(c_i^\dagger\) and \(c_i\) are the electron creation and annihilation operators at site \(i\), with \(c_i=(c_{i\uparrow},c_{i\downarrow})^T\); \(\langle ij\rangle\) denotes the nearest-neighbor site pairs included in the model; \(t_{ij}^{\mathrm{layer}}\) is the coupling between the two molecular layers; and \(t_{ij}^{\mathrm{up}},t_{ij}^{\mathrm{dn}}\) are the nearest-neighbor hoppings in the upper and lower molecules, respectively.

The coupling between the molecule and the upper and lower electrodes is
\[
\mathcal{H}_{\mathrm{el}}
=\sum_{\alpha=L,U}\left(c_D^\dagger V_{D\alpha}a_\alpha+\mathrm{H.c.}\right), \tag{23}
\]
where \(c_D=(c_i)_{i\in D}\) is the column vector of site operators in the central molecular region, \(a_\alpha=(a_{k\alpha})\) is the column vector of electron annihilation operators for mode \(k\) in electrode \(\alpha\), and \(V_{D\alpha}\) is the coupling matrix between the molecule and electrode \(\alpha\). The coupling between the molecule and environmental degrees of freedom is
\[
\mathcal{H}_{\mathrm{d}}
=\left(c_D^\dagger V_d b+\mathrm{H.c.}\right), \tag{24}
\]
where \(b=(b_q)\) is the column vector of annihilation operators for the environmental modes and \(V_d\) is the coupling matrix between the molecule and these modes.

Within the nonequilibrium Green's function framework, the electrode and environmental degrees of freedom are integrated out, and their effects on the central molecular region are described by retarded self-energies:
\[
\Sigma_\alpha^r(E)=V_{D\alpha}g_\alpha^r(E)V_{D\alpha}^\dagger,\quad \alpha=L,U, \tag{25}
\]
\[
\Sigma_d^r(E)=V_d g_d^r(E)V_d^\dagger, \tag{26}
\]
where \(g_\alpha^r\) is the surface retarded Green's function of the isolated electrode \(\alpha\), and \(g_d^r\) is the retarded Green's function of the environmental modes. In the wide-band approximation,
\[
\Sigma_d^r(E)\simeq-\frac{i}{2}\Gamma_d I, \tag{27}
\]
with \(\Gamma_d\) the tunable standard leakage broadening. The retarded self-energies of the upper and lower electrodes are taken as
\[
\Sigma_\alpha^r=-\frac{i}{2}\Gamma_\alpha, \tag{28}
\]
where
\[
\Gamma_\alpha=\gamma_\alpha C_\alpha\otimes I_2,\quad \gamma_\alpha=1~\mathrm{eV}, \tag{29}
\]
\(\gamma_\alpha\) is the coupling broadening energy of electrode \(\alpha\), \(C_\alpha\) is the dimensionless contact matrix of the corresponding electrode, the number of active channels excluding spin is \(M_\alpha=\mathrm{rank}\,C_\alpha\), and \(I_2\) is the identity matrix in spin space. These quantities satisfy
\[
\Gamma_\alpha(E)=i\left[\Sigma_\alpha^r(E)-\Sigma_\alpha^a(E)\right]. \tag{30}
\]
The retarded Green's function of the central molecular region is
\[
G^r(E)=\left[EI-H_D-\Sigma_L^r-\Sigma_U^r-\Sigma_d^r\right]^{-1}, \tag{31}
\]
and the advanced Green's function is \(G^a(E)=[G^r(E)]^\dagger\), where \(H_D=\mathcal{H}_{\mathrm{mol}}+\mathcal{H}_{\mathrm{geo}}\) is the Hamiltonian of the central molecular region.

\subsection*{Transport formulation}

We calculate the transport properties within the nonequilibrium Green's function framework. For unpolarized electrons incident from the \(L\) electrode and outgoing from the \(U\) electrode, the spin-resolved transmission coefficient is
\[
T_\sigma(E)=\mathrm{Tr}\left[\Gamma_{U,\sigma} G^r(E)\Gamma_LG^a(E)\right], \tag{32}
\]
where \(\Gamma_L\) is the broadening matrix of the \(L\) electrode, \(\Gamma_{U,\sigma}=P_\sigma\Gamma_U P_\sigma\), with \(P_\sigma\) the projection operator for spin \(\sigma\), and \(G^r(E)\) and \(G^a(E)\) are the retarded and advanced Green's functions of the central molecular region, respectively. The transmission difference between the two spin channels is defined as
\[
\Delta T=T_\uparrow-T_\downarrow. \tag{33}
\]
From \(T_\sigma(E)\), the spin-resolved current is obtained as
\[
I_\sigma=\frac{e}{h}\int dE\,T_\sigma(E)\left[f_L(E)-f_U(E)\right], \tag{34}
\]
where \(f_{L,U}(E)\) are the Fermi distribution functions of the left and right electrodes. The spin polarization of the transmitted current contributed by the two observed terminals is then defined as
\[
SP=\frac{I_\uparrow-I_\downarrow}{I_\uparrow+I_\downarrow}. \tag{35}
\]
These quantities, \(\Delta T\), \(I_\uparrow\), \(I_\downarrow\), and \(SP\), are used in the main text to characterize the spin-dependent transport. Situations that do not satisfy the protection conditions of the classical and extended no-go theorems allow a nonzero \(SP\) to emerge, which is precisely the regime studied by the CISS effect.

\end{document}